%% file: main_Md.tex
\documentclass[final,5p,times,twocolumn]{elsarticle}
\biboptions{sort&compress}
\usepackage{graphicx}
\usepackage{amsmath}
\usepackage{amssymb}
\usepackage{bm}
\usepackage{comment}
\usepackage{tabularx}
\usepackage[hidelinks]{hyperref}

\journal{Physics Letters B}

\begin{document}
\begin{frontmatter}
\title{Fission Modes and Fragment Shell Structures in $^{258}$Md$^*$ from Six-Dimensional Langevin Calculations}

\author[jaea]{Kazuki Okada\corref{cor1}}
\ead{okada.kazuki@jaea.go.jp}
\cortext[cor1]{Corresponding author}
\author[jaea]{Katsuhisa Nishio}
\author[kansai]{Takahiro Wada}
\author[jinr,bordeaux]{Nicolae Carjan}
\affiliation[jaea]{organization={Advanced Science Research Center, Japan Atomic Energy Agency}, city={Tokai}, state={Ibaraki}, postcode={319-1195}, country={Japan}}
\affiliation[kansai]{organization={Department of Pure and Applied Physics, Kansai University}, city={Suita}, state={Osaka}, postcode={564-8680}, country={Japan}}
\affiliation[jinr]{organization={Joint Institute for Nuclear Research}, city={Dubna}, postcode={141980}, country={Russia}}
\affiliation[bordeaux]{organization={University of Bordeaux, CNRS, LP2i Bordeaux, UMR 5797}, city={Gradignan}, postcode={F-33170}, country={France}}

\begin{abstract}
The fission of $^{258}$Md$^*$ is calculated in the excitation energy range of $E^*=6$--36~MeV using a six-dimensional Langevin framework with the Cassini shape parametrization.
The calculated events are classified into two symmetric and two asymmetric fission modes based on the fragment mass and the quadrupole deformations of the two fragments at scission.
The symmetric modes are the short mode with high total kinetic energy (TKE) and the superlong mode with low TKE, whereas the asymmetric modes differ in mass asymmetry.
With increasing excitation energy, the yield of the short mode decreases, whereas the combined yield of the two asymmetric modes increases, as observed in the in-beam prompt-fission study of $^{258}$Md$^*$.
From an analysis of the fragment shapes and associated single-particle levels, the short mode and the dominant asymmetric mode with the smaller mass asymmetry are found to involve a compact fragment characterized by deformed shell gaps at $Z=52$ and $N=84$, while the complementary fragment is compact in the short mode and strongly elongated in the asymmetric mode.
\end{abstract}

\begin{keyword}
Nuclear fission \sep Fission modes \sep Langevin equation \sep Cassini shape parametrization \sep Fragment shell structure
\end{keyword}
\end{frontmatter}

\section{Introduction}
\input{intro_Md}

\section{Mode classification}
\input{method_Md}

\section{Results and discussion}
\input{result_Md}

\input{result2_Md}

\section{Summary and conclusions}

In summary, the fission of $^{258}$Md$^*$ was calculated over $E^*=6$--36~MeV using a six-dimensional Langevin framework based on the Cassini shape parametrization.
The calculated events were classified into four modes, SH, AS1, AS2, and SL, based on the fragment mass and on the quadrupole deformations of both fragments at scission.
SH and SL are symmetric modes with high and low TKE, respectively.
AS2 is more mass-asymmetric than AS1. SL has only a very low yield.
This classification was used to analyze the excitation energy dependence of the event fraction, average TKE, and mean heavy fragment mass for each mode, and the single-particle levels at the scission point were determined for each mode.

At $E^*=6$~MeV, the high-TKE SH mode has the largest individual mode yield, accounting for nearly 50\% of the events, with smaller contributions from AS1 and AS2.
With increasing $E^*$, the event fraction of SH decreases rapidly, whereas the combined event fraction of AS1 and AS2 increases.
The contribution from SL remains small throughout the calculated range.
At the same time, the average TKE generally decreases, with the most pronounced decrease for SH.
These changes with excitation energy are attributed to shell damping.
At the highest excitation energy, the mode distributions overlap more strongly and form a broad distribution over the mass--TKE plane (Supplementary Fig.~S1).

The fragment shapes at scission and the corresponding single-particle levels provide insight into the nuclear-structure origin of each mode.
The compact fragments in SH and the compact AS1 heavy fragment show similar deformations and share deformed shell gaps at $Z=52$ and $N=84$.
Despite this similarity, the complementary fragments in SH and AS1 have different masses and deformations.
The AS2 scission configuration is characterized by both fragments being appreciably deformed and by the light fragment showing a proton shell gap at $Z=38$.
SL does not show evident shell gaps, indicating a more liquid-drop-like character at scission.

From a comparison with the results for $^{236}$U fission in Ref.~\cite{Okada236U_companion}, we find that similar fragment shell patterns can occur in distinct mode configurations.
The SH fragments and the AS1 heavy fragment in $^{258}$Md$^*$ are stabilized by deformed shell gaps, as is the heavy fragment of the dominant asymmetric mode in $^{236}$U.
This mechanism distinguishes SH from SS, whose spherical fragments are stabilized by the shell closures of $^{132}$Sn.
The elongated AS1 light fragment in $^{258}$Md$^*$ exhibits a shell pattern similar to that of the symmetric SL mode in $^{236}$U.

\section*{Data availability}
Data will be made available on request.

\section*{Acknowledgements}
This work was supported by JSPS KAKENHI Grant Number JP24K22887.
The six-dimensional Langevin calculations and mode classification were performed on the JAEA supercomputer HPE SGI8600.

\bibliographystyle{elsarticle-num}
\bibliography{myref_Md}

\end{document}


\begin{center}
{\large\bfseries Supplementary Material}\par\vspace{2pt}
{\large\bfseries Fission Modes and Fragment Shell Structures in $^{258}$Md$^*$ from Six-Dimensional Langevin Calculations}\par\vspace{2pt}
Kazuki Okada, Katsuhisa Nishio, Takahiro Wada, and Nicolae Carjan
\end{center}

\section{Six-dimensional Langevin calculations}

The fission dynamics of $^{258}$Md$^*$ is calculated using the six-dimensional Langevin framework.
The nuclear shape is specified by the Cassini coordinates~\cite{pashkevich1971}
$\bm q=(\alpha,\alpha_1,\alpha_2,\alpha_3,\alpha_4,\alpha_5)$.
Here, $\alpha$ describes elongation, while the coefficients $\alpha_1,\ldots,\alpha_5$ introduce additional symmetric and asymmetric deformations.

The collective coordinates $q_i$ and their conjugate momenta $p_i$ evolve according to
\begin{align}
    \frac{dq_i}{dt}
    &= (m^{-1})_{ij}p_j,
    \nonumber\\
    \frac{dp_i}{dt}
    &= -\left.\frac{\partial F}{\partial q_i}\right|_T
       -\frac{1}{2}\frac{\partial(m^{-1})_{jk}}{\partial q_i}p_jp_k
       -\gamma_{ij}(m^{-1})_{jk}p_k
       +g_{ij}R_j(t).
    \label{eq:supp_langevin}
\end{align}
Here, $F(\bm q,T)$ is the free energy, and $m_{ij}(\bm q,T)$ and $\gamma_{ij}(\bm q,T)$ are the collective inertia and friction tensors.
Both tensors are calculated microscopically within linear response theory~\cite{ivanyuk1999}.
The noise terms $R_i(t)$ satisfy $\langle R_i(t)\rangle=0$ and
$\langle R_i(t)R_j(t')\rangle=2\delta_{ij}\delta(t-t')$.
The coefficients $g_{ij}$ determine the random-force strength and satisfy the modified Einstein relation $\sum_k g_{ik}g_{jk}=T_{\mathrm{eff}}\gamma_{ij}$, where $T_{\mathrm{eff}}$ is the effective temperature calculated as in~\cite{hofmann1979,usang2017}.

The free energy is evaluated in the macroscopic--microscopic approach as
\begin{equation}
    F(\bm q,T)
    = V_{\mathrm{FRLDM}}(\bm q)
      -a(\bm q)T^2
      +F_{\mathrm{micro}}(\bm q,T).
    \label{eq:supp_free_energy}
\end{equation}
Here, $V_{\mathrm{FRLDM}}$ is the finite-range liquid-drop potential~\cite{moller2016}, and $a(\bm q)$ is the level density parameter~\cite{toke1981}.
The microscopic term includes shell corrections evaluated using the Strutinsky method~\cite{strutinsky1967,strutinsky1968} and pairing corrections evaluated within the BCS approximation at finite temperature~\cite{ivanyuk2018}.

The intrinsic excitation energy $E_{\mathrm{int}}$ is updated at each time step from the condition of energy conservation,
\begin{equation}
    E_{\mathrm{int}}
    = E_{\mathrm{tot}}
      -\frac{1}{2}(m^{-1})_{ij}p_ip_j
      -F(\bm q,0),
    \label{eq:supp_temperature}
\end{equation}
where $E_{\mathrm{tot}}$ is the total energy.
The nuclear temperature $T$ is determined from the Fermi-gas relation $E_{\mathrm{int}}=a(\bm q)T^2$.

The Langevin trajectories are started from the ground-state configuration of $^{258}$Md.
A scission configuration is identified when the neck radius becomes $2~\mathrm{fm}$ or smaller.
Approximately $3\times10^5$ scission events are analyzed at each excitation energy.
The total kinetic energy (TKE) is evaluated at scission as the sum of the Coulomb repulsion energy between the two nascent fragments and their prescission kinetic energy.

\section{Mode classification}

The $k$-means classification follows the procedure used for $^{236}$U~\cite{Okada236U_companion} and is performed independently at each excitation energy.
Each fragment $F$ is represented by $\bm y_n=(A_F,\beta_{2,F},\beta_{2,\bar F})$, where $\bar F$ denotes its partner and $n$ indexes the registered fragments.
Here, $A_F$ and $\beta_{2,F}$ are the mass number and quadrupole deformation parameter of fragment $F$, respectively.
Both fragments from each scission event are included.
The three variables are standardized as
\begin{equation}
    x_{n,i}=\frac{y_{n,i}-\bar y_i}{s_i},
    \qquad i=1,2,3,
    \label{eq:supp_standardization}
\end{equation}
where $\bar y_i$ and $s_i$ are the mean and standard deviation of the $i$th variable over all registered fragments at the given excitation energy.

The number of numerical clusters is set to $K=6$: two mirror clusters for each of AS1 and AS2, and one cluster for each of SH and SL.
The standardized points $\bm x_n$ are partitioned into clusters by iteratively reducing the within-cluster sum of squared Euclidean distances,
\begin{equation}
    W=\sum_{c=1}^{K}\sum_{n\in C_c}
    \left\|\bm x_n-\bm\mu_c\right\|^2,
    \label{eq:supp_kmeans_objective}
\end{equation}
where $C_c$ is the set of indices of points assigned to cluster $c$ and $\bm\mu_c$ is its centroid.
Each point is assigned to the nearest centroid, and each centroid is updated to the mean of its assigned points:
\begin{align}
    c(n)&=\arg\min_c\left\|\bm x_n-\bm\mu_c\right\|^2,
    \nonumber\\
    \bm\mu_c&=\frac{1}{N_c}\sum_{n\in C_c}\bm x_n.
    \label{eq:supp_kmeans_update}
\end{align}
Here, $N_c$ is the number of points in cluster $c$.
The assignment and centroid update steps are repeated until convergence.

After convergence, the cluster pairs related by exchange of the two fragments are combined for the asymmetric modes.
The resulting groups, together with the symmetric components, are assigned to SH, AS1, AS2, and SL according to their fragment masses and deformations.
Each fragment was assigned to the same mode for all initial centroid configurations examined.

\section{Mode-resolved mass--TKE distributions}

\begin{figure}[!ht]
    \centering
    \includegraphics[width=0.32\textwidth,trim={39bp 40bp 0bp 152bp},clip]{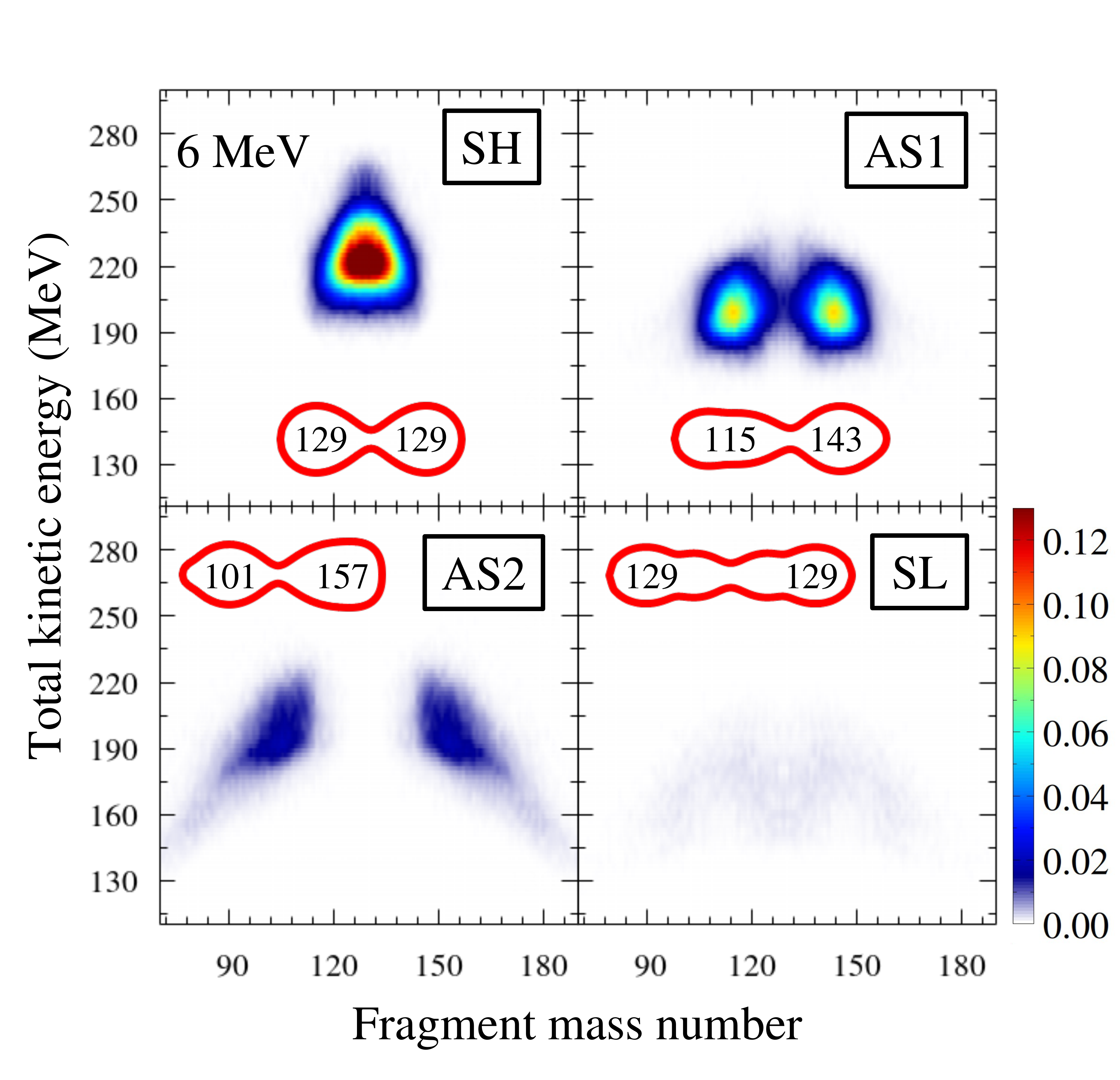}
    \hfill
    \includegraphics[width=0.32\textwidth,trim={39bp 40bp 0bp 152bp},clip]{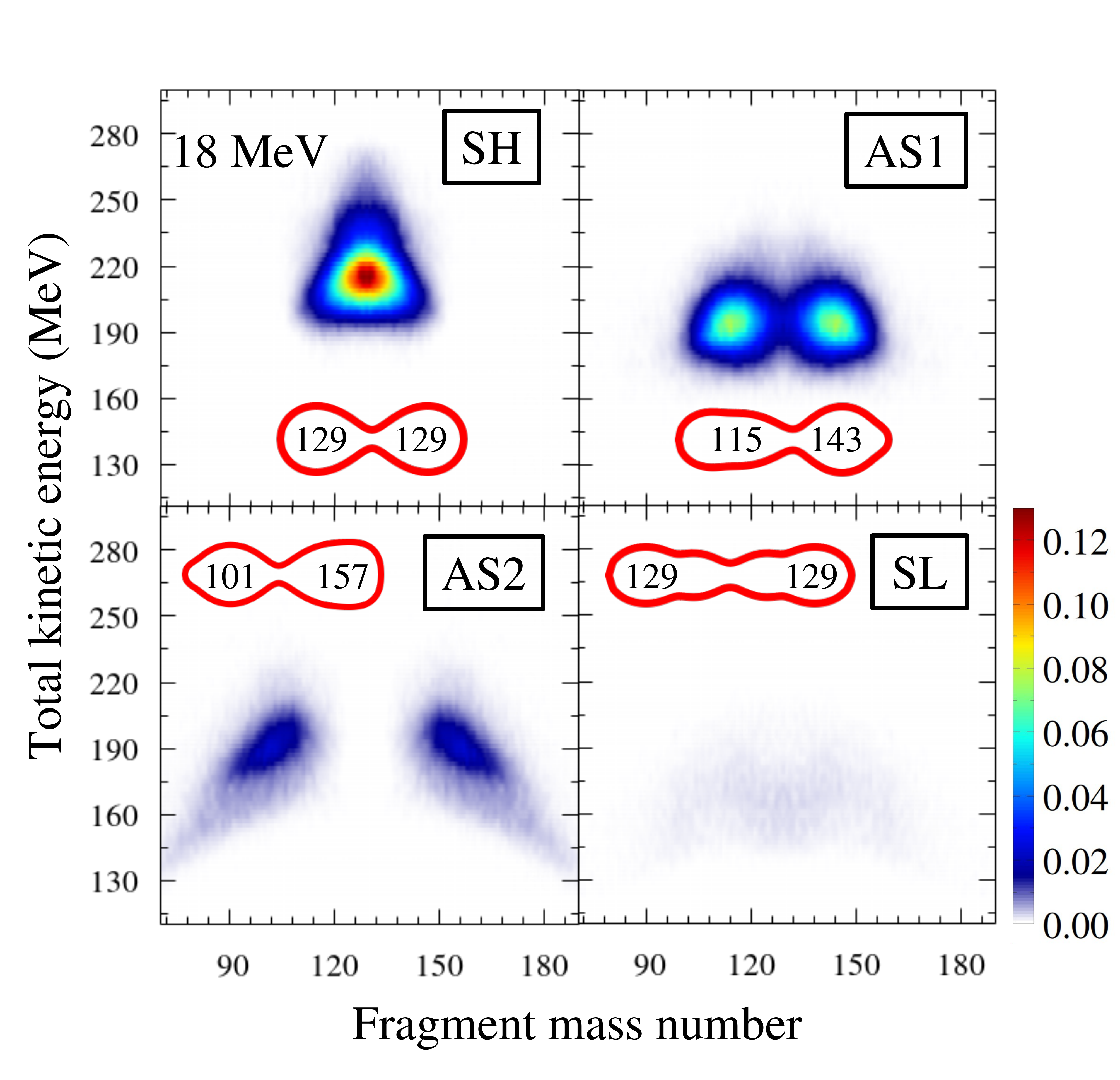}
    \hfill
    \includegraphics[width=0.32\textwidth,trim={39bp 40bp 0bp 152bp},clip]{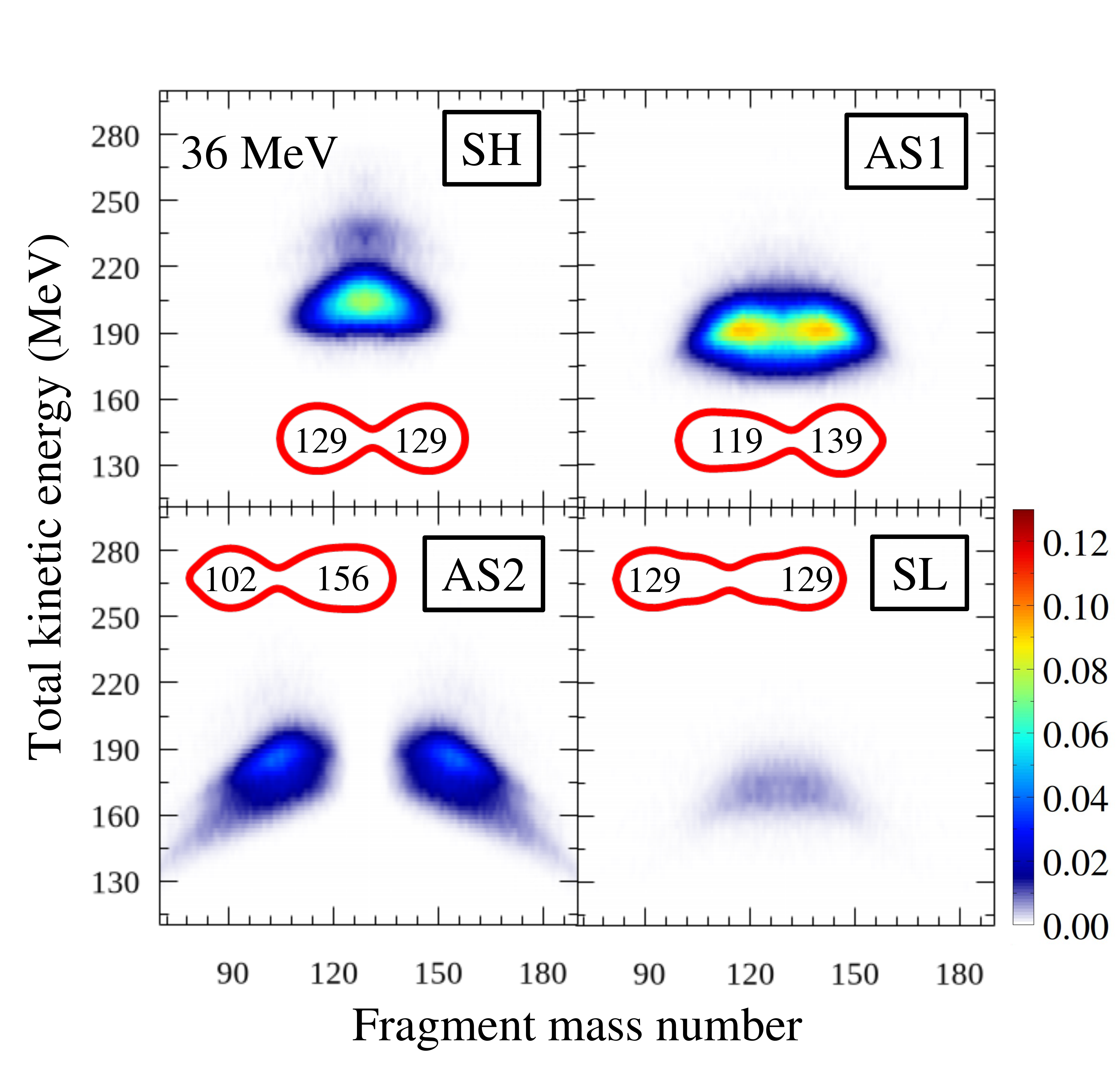}
    \caption{Fragment mass--TKE distributions for the SH, AS1, AS2, and SL modes in $^{258}$Md$^*$ fission.
        The three sets of panels from left to right correspond to $E^*=6$, 18, and 36~MeV.
        Within each set, the upper left, upper right, lower left, and lower right panels show SH, AS1, AS2, and SL, respectively.
        The sum of the four distributions at each energy gives the total distribution shown in Fig.~2 of the main article.
        Representative scission shapes are also displayed with the corresponding fragment mass numbers.}
    \label{fig:mode_tke}
\end{figure}

Supplementary Fig.~\ref{fig:mode_tke} shows the fragment mass--TKE distributions for each mode at $E^*=6$, 18, and 36~MeV.
At $E^*=6$~MeV, SH is concentrated near mass symmetry at a high TKE of approximately $220$~MeV.
AS1 is confined to localized regions of the mass--TKE plane.
AS2 is located in a more asymmetric mass region than AS1 and shows oblique distributions with tails extending toward lower TKE.
Only a very small SL component appears near mass symmetry at low TKE.

At $E^*=18$~MeV, the total distribution in Fig.~2 of the main article extends over a broad asymmetric mass region, as observed in the experimental mass--TKE distribution~\cite{nishio2025}.
Within the present shape classification, the neighboring regions occupied by AS1 and AS2 partially overlap in the mass--TKE plane.
Consequently, their mass projections merge into a single broad asymmetric distribution.
At $E^*=36$~MeV, a substantial shift of SH toward lower TKE is observed together with an overall broadening of the mass--TKE distributions.
The classification was also attempted at excitation energies above $36$~MeV, but the shell signatures that characterize the individual modes at lower energies could no longer be identified.

\begingroup
\small
\bibliographystyle{elsarticle-num}
\bibliography{myref_Md}
\endgroup

%% file: intro_Md.tex
Nuclear fission is a fundamental process in heavy nuclei, and predictive descriptions
of fission are required for understanding its role in $r$-process
nucleosynthesis~\cite{giuliani2020,wu2019,eichler2015,goriely2015,mumpower2018,vassh2019}.
In the fission of actinides at low excitation energy, fragment mass and total kinetic
energy (TKE) distributions are commonly interpreted in terms of fission modes,
i.e., the pathways on the potential energy surface along which the shape of the fissioning nucleus evolves.
Experimentally observed asymmetric fission (AS) is commonly described in terms of the Standard I and Standard II modes,
whereas the superlong (SL) mode appears as a symmetric component at lower TKE~\cite{brosa1990}.
The observation of these modes indicates that fragment mass and kinetic energy contain information about
different scission configurations.
In particular, microscopic calculations have shown that deformed shell gaps around
$Z=52,56$ and $N=84,88$, together with octupole deformation in the $^{144}$Ba region,
play an important role in asymmetric fission of actinides~\cite{scamps2018}.

In the spontaneous fission of heavier actinides, the measured fragment mass distribution changes from an asymmetric
distribution in $^{256}$Fm to a narrow symmetric distribution in $^{258}$Fm~\cite{hulet1986,hulet1989}.
This transition is accompanied by the appearance of a symmetric component with high TKE,
indicating compact scission configurations.
A systematic study of spontaneous fission has shown that the symmetric component becomes prominent in fissioning
nuclei around $A\simeq257$ and above~\cite{hoffman1995}.
The sharp symmetric fission mode is called supershort (SS) and has been identified in spontaneous fission of $^{258,259}$Fm, $^{259,260}$Md, and $^{262}$No. It is associated with shell structure near $^{132}$Sn~\cite{hulet1986,hulet1989,wild1990,lougheed1989,hulet1980}.

In the region of compound nuclei with $A_c>257$, fission of $^{258}$Md$^*$ was measured at the Japan Atomic Energy Agency~\cite{nishio2025}.
The compound nucleus was produced by bombarding a short-lived $^{254}$Es target with a
$^4$He beam.
At an excitation energy of $E^*=15$~MeV, the measured mass--TKE distribution shows comparable contributions
from asymmetric and symmetric fission.
The symmetric component is referred to as the short (SH) mode; it has a TKE of about
$210$~MeV, lower than that of the SS mode but higher than that of the AS mode.
When the excitation energy is increased to $E^*=18$~MeV, the asymmetric contribution
increases, in contrast to the trend in the fission of lighter actinides.
This change over just a 3~MeV interval indicates that the competition between SH and
AS in $^{258}$Md$^*$ is highly sensitive to excitation energy.
A weak component with low TKE, compatible with an SL mode, was also suggested by the
measurement.
In this context, the excited compound nucleus $^{258}$Md$^*$ is a useful system for
studying different fission modes and their excitation energy dependence.

A dynamical model based on Langevin equations has the advantage of directly generating
fragment mass and TKE distributions, thereby allowing comparison with experimental
data~\cite{wada1993,karpov2001,aritomo2013,sierk2017,ishizuka2017,usang2017,liu2021,ivanyuk2025,okada2025}.
Shape parametrizations based on the two-center shell model have been widely used in Langevin calculations~\cite{aritomo2013,ishizuka2017,ivanyuk2024,ivanyuk2025}.
In scission-point models, the Cassini parametrization is well suited to describing fragment shapes at scission through the inclusion of higher-order shape degrees of freedom~\cite{pashkevich1971,carjan2015,carjan2019,ivanyukCarjan2024}.
It has recently been applied in a five-dimensional Langevin calculation~\cite{okada2025}.
The fragment shapes obtained from calculations using such a multidimensional shape parametrization can be used to examine single-particle levels and shell gaps.

In this work, we investigate the fission of $^{258}$Md$^*$ over the excitation energy range $E^*=6$--36~MeV using a six-dimensional Langevin calculation based on the Cassini shape parametrization.
Following the mode classification used in a six-dimensional Langevin study of $^{236}$U fission~\cite{Okada236U_companion}, we classify the calculated events on the basis of the fragment mass and the quadrupole deformations of both fragments at scission.
For each mode, we examine the excitation energy dependence of the event fraction, average TKE, and mean heavy fragment mass.
We further analyze representative fragment shapes and single-particle levels to clarify the role of shell structure in each fission mode.

%% file: method_Md.tex
The fission dynamics is described by the six-dimensional Langevin framework.
The nuclear shape is specified by the Cassini coordinates $\bm q=(\alpha,\alpha_1,\alpha_2,\alpha_3,\alpha_4,\alpha_5)$, where $\alpha$ describes elongation and the remaining coordinates introduce additional symmetric and asymmetric deformations.
The dynamical equations are summarized in the Supplementary Material.

In this work, the mass numbers and quadrupole deformations of both fragments are considered together to classify fission events.
The classification variables are $(A_F,\beta_{2,F},\beta_{2,\bar F})$, where $A$ is the mass number and $\beta_2$ the quadrupole deformation parameter.
The subscripts $F$ and $\bar F$ denote a fragment and its partner, respectively.
$A_{\bar F}$ is omitted because it is automatically determined by $A_F$.
TKE and the octupole deformation $\beta_3$ are not used in the classification but are evaluated after the mode assignment.

The classification is performed independently at each excitation energy using the $k$-means algorithm~\cite{macqueen1967,lloyd1982}, with the light and heavy fragment sides treated symmetrically.
The cluster pairs related by exchange of the light and heavy fragments are combined for the asymmetric modes.
The combined asymmetric components and the symmetric components are assigned to four physical modes: SH, AS1, AS2, and SL.
SH and SL correspond to compact and elongated symmetric configurations, respectively.
AS1 and AS2 are asymmetric modes with different fragment masses and deformations, with AS2 having the larger mass asymmetry.
The labels AS1 and AS2 distinguish the present classification from the conventional Standard I and Standard II modes.
Details of the classification procedure are given in the Supplementary Material.

For each mode, a representative scission configuration is constructed from the mean Cassini parameters of the assigned events.
The proton and neutron numbers of the representative fragments are assigned from their mean masses using the $N/Z$ ratio of $^{258}$Md.
Single-particle levels are then calculated for the isolated fragments with these shapes and particle numbers to examine the shell structures associated with each mode.

%% file: result_Md.tex
\begin{figure}[t]
    \centering
    \includegraphics[width=\columnwidth,trim={39bp 41bp 41bp 44bp},clip]{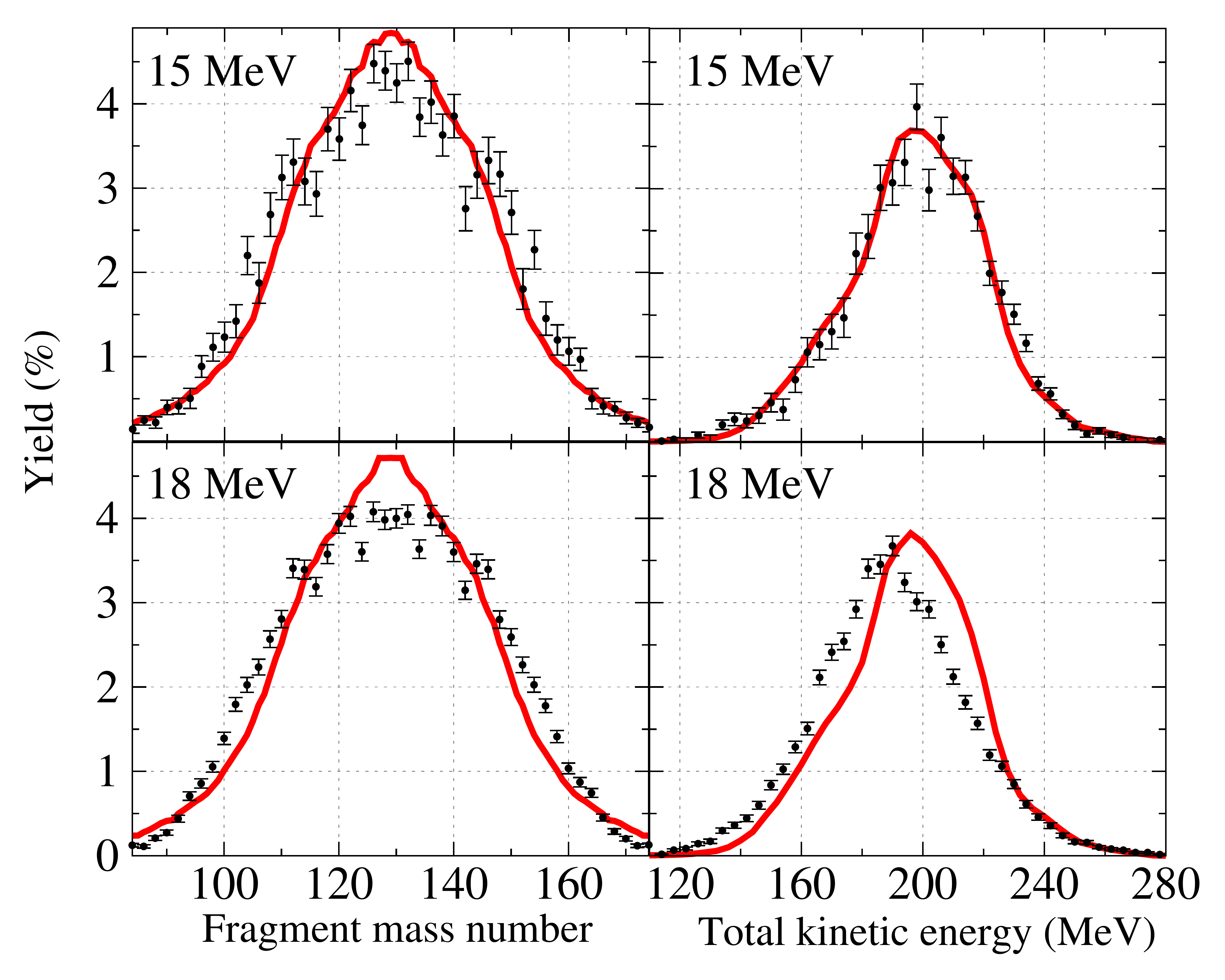}
\caption{Fragment mass distributions in the left panels and TKE distributions in the right panels for $^{258}$Md$^*$ fission at $E^*=15$ and $18$~MeV.
The red curves and black circles denote the calculated results and the experimental data given in~\cite{nishio2025}, respectively.}
    \label{fig:ex}
\end{figure}

The calculated fragment mass and TKE distributions at $E^*=15$ and $18$~MeV are compared with the measured distributions in Fig.~\ref{fig:ex}.
At both excitation energies, the fragment mass and TKE distributions are single-peaked.
The calculation reproduces the overall shapes of the mass and TKE distributions at both excitation energies.
A comparison of the calculated two-dimensional mass--TKE distributions with the experimental data is presented in Ref.~\cite{nishio2025}.

\begin{figure*}[t]
    \centering
    \includegraphics[width=\textwidth,trim={23bp 33bp 39bp 21bp},clip]{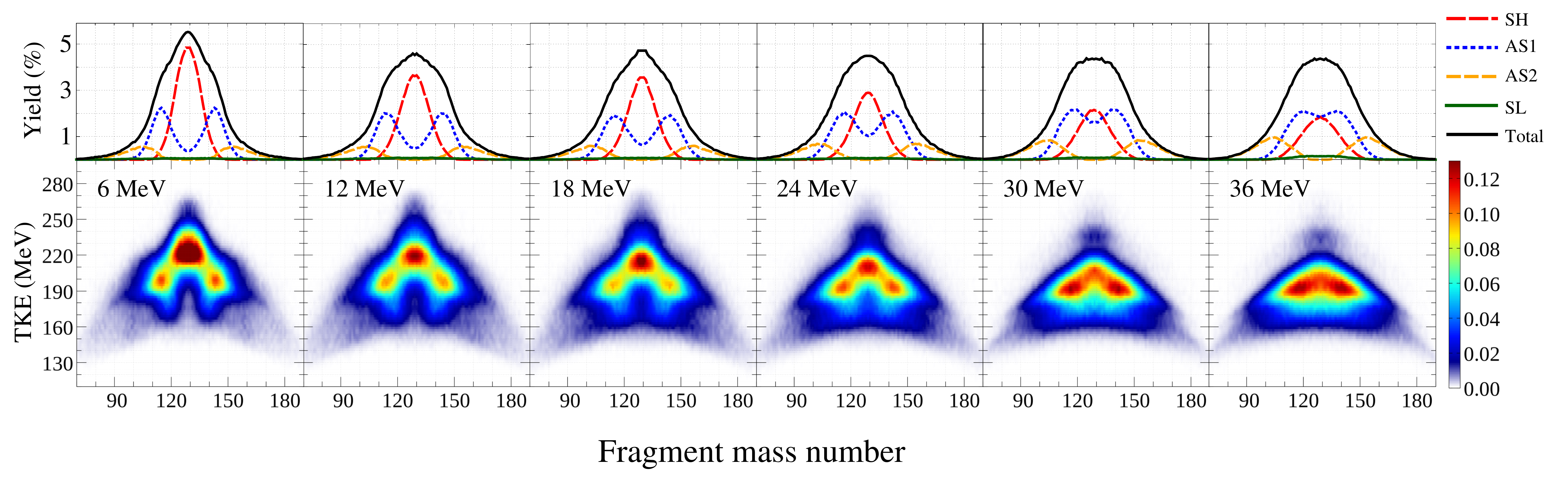}
\caption{Fragment mass--TKE distributions (lower panels) and the corresponding fragment mass distributions (upper panels) for $^{258}$Md$^*$ fission at $E^*=6$--36~MeV.
In the lower panels, each distribution is normalized to a total yield of 200\%, with bin widths of 1 mass unit for fragment mass and 4~MeV for TKE.
The colored curves in the upper panels show the fragment mass distributions for the SH, AS1, AS2, and SL modes, whereas the black curve shows the total fragment mass distribution.}
    \label{fig:TKEFMD}
\end{figure*}

To examine the excitation energy dependence of the fragment mass and TKE distributions for the individual fission modes, the Langevin calculations are performed over $E^*=6$--36~MeV, and the modes are identified from their scission configurations.
The resulting mass--TKE distributions are shown in the lower panels of Fig.~\ref{fig:TKEFMD}, while the projected mass distributions are shown by the black curves in the upper panels.
At low excitation energies, the high-TKE symmetric component, which we assign to the SH mode, has the largest individual mode yield, although substantial asymmetric yields are also found.
The high-TKE symmetric yield decreases with $E^*$ while the asymmetric yield increases.
With increasing excitation energy, the SH distribution shifts toward lower TKE and increasingly overlaps with the asymmetric component, which consists of the two asymmetric modes (AS1 and AS2) in the analysis below.
At $E^*=36$~MeV, the SH and asymmetric components are no longer clearly resolved in the mass--TKE plane.

The mass--TKE distributions for each mode are presented individually for $E^*=6$, 18, and 36~MeV in Supplementary Fig.~S1.
To visualize the excitation energy dependence of the mode contributions, the colored curves in the upper panels of Fig.~\ref{fig:TKEFMD} show the fragment mass distributions for each mode.
Over most of the calculated excitation energy range, SH and AS1 together make the dominant contribution to the total fragment yield.
AS2 exhibits a larger mass asymmetry than AS1.
The SL contribution appears only as a peak with very low intensity.

%% file: result2_Md.tex
\begin{figure}[t]
	\centering
	\includegraphics[width=\columnwidth,trim={27bp 29bp 36bp 39bp},clip]{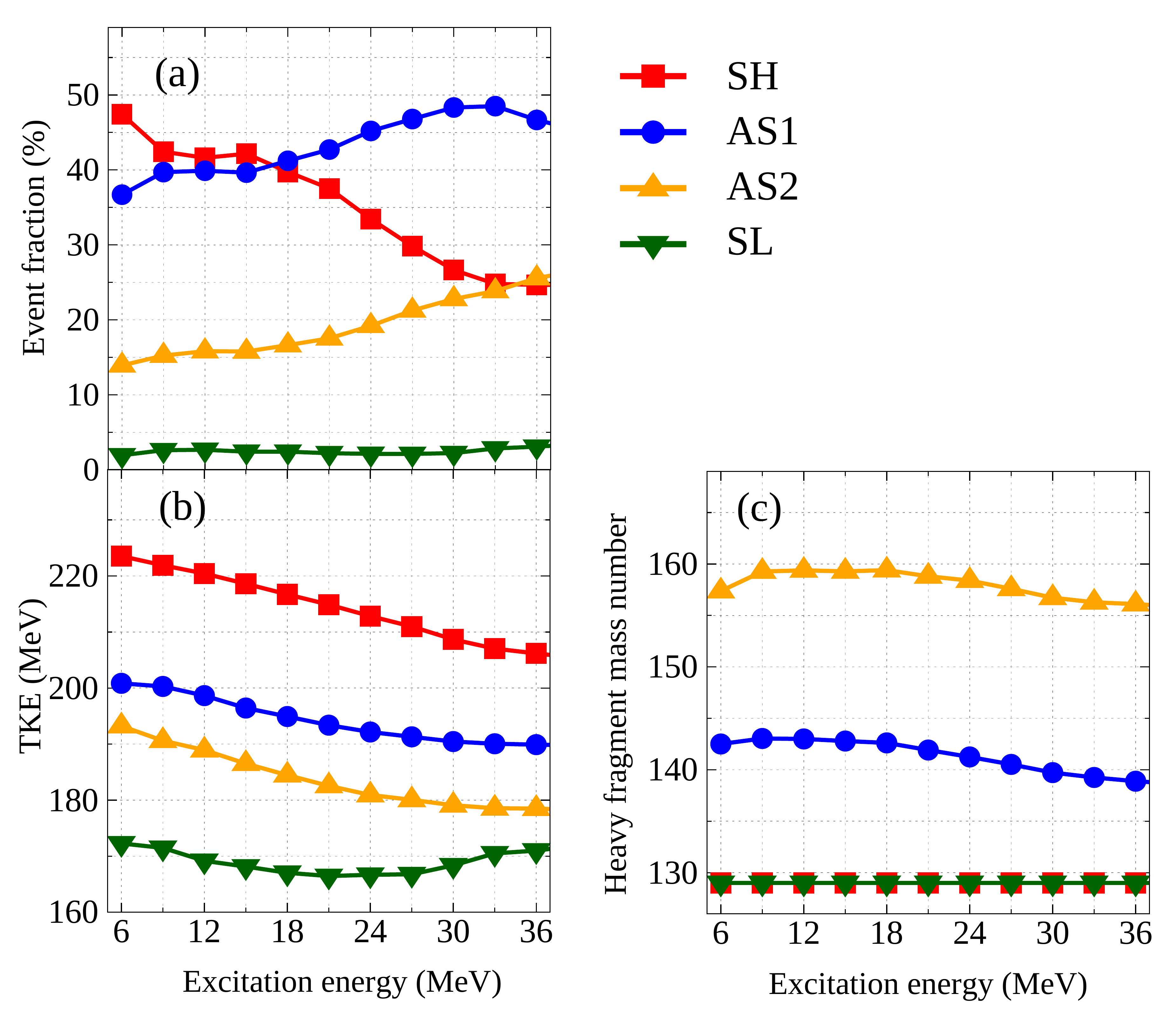}
	\caption{Excitation energy dependence of (a) the event fraction, (b) the average TKE, and (c) the mean heavy fragment mass for the SH, AS1, AS2, and SL modes in $^{258}$Md$^*$ fission.
		The red, blue, orange, and green lines denote SH, AS1, AS2, and SL, respectively.}
	\label{fig:yield_tke}
\end{figure}

Figure~\ref{fig:yield_tke} shows the excitation energy dependence of the event fraction, average TKE, and mean heavy fragment mass for each mode.
Figure~\ref{fig:yield_tke}(a) quantifies the excitation energy dependence of the mode contributions shown in the upper panels of Fig.~\ref{fig:TKEFMD}.
At $E^*=6$~MeV, the event fraction of SH is nearly 50\%.
With increasing excitation energy, the event fraction of SH decreases rapidly, whereas those of both asymmetric modes, AS1 and AS2, increase, consistent with the experimentally observed increase in the combined asymmetric contribution~\cite{nishio2025}.
The contribution from SL remains small throughout the calculated excitation energy range.
This small contribution is also consistent with the experimental data~\cite{nishio2025}.

Figure~\ref{fig:yield_tke}(b) shows the average TKE for all the modes.
SH has the highest average TKE throughout the excitation energy range, whereas SL has the lowest.
The average TKEs of AS1 and AS2 are intermediate between those of SH and SL, with AS2 having the lower value.
The average TKE generally decreases with increasing $E^*$, with the most pronounced decrease for SH.
In terms of the potential energy surface, this trend suggests that the scission configurations assigned to SH gradually lose their compact character as shell effects are damped.

The mean heavy fragment masses of the four modes are shown in Fig.~\ref{fig:yield_tke}(c).
The symmetry $A_H=A_L=129$ of the SH and SL modes is confirmed over the full excitation-energy range.
The mean heavy fragment masses of the asymmetric modes show a slight shift toward mass symmetry with increasing $E^*$.
This shift can be understood as a consequence of shell damping, which progressively reduces the stabilizing influence of shell effects on asymmetric configurations.
In the mode analysis of the experimental data~\cite{nishio2025}, the centroid of the asymmetric component shifted toward mass symmetry from $E^*=15$ to 18~MeV, and the average TKE of the asymmetric mode decreased with excitation energy. Both trends are consistent with the present calculation. In contrast, the excitation energy dependence of the average TKE of the SH mode could not be resolved experimentally because of the strong overlap between the SH and asymmetric-mode yields.

\begin{figure}[t]
	\centering
	\includegraphics[width=\columnwidth,trim={147bp 33bp 67bp 0bp},clip]{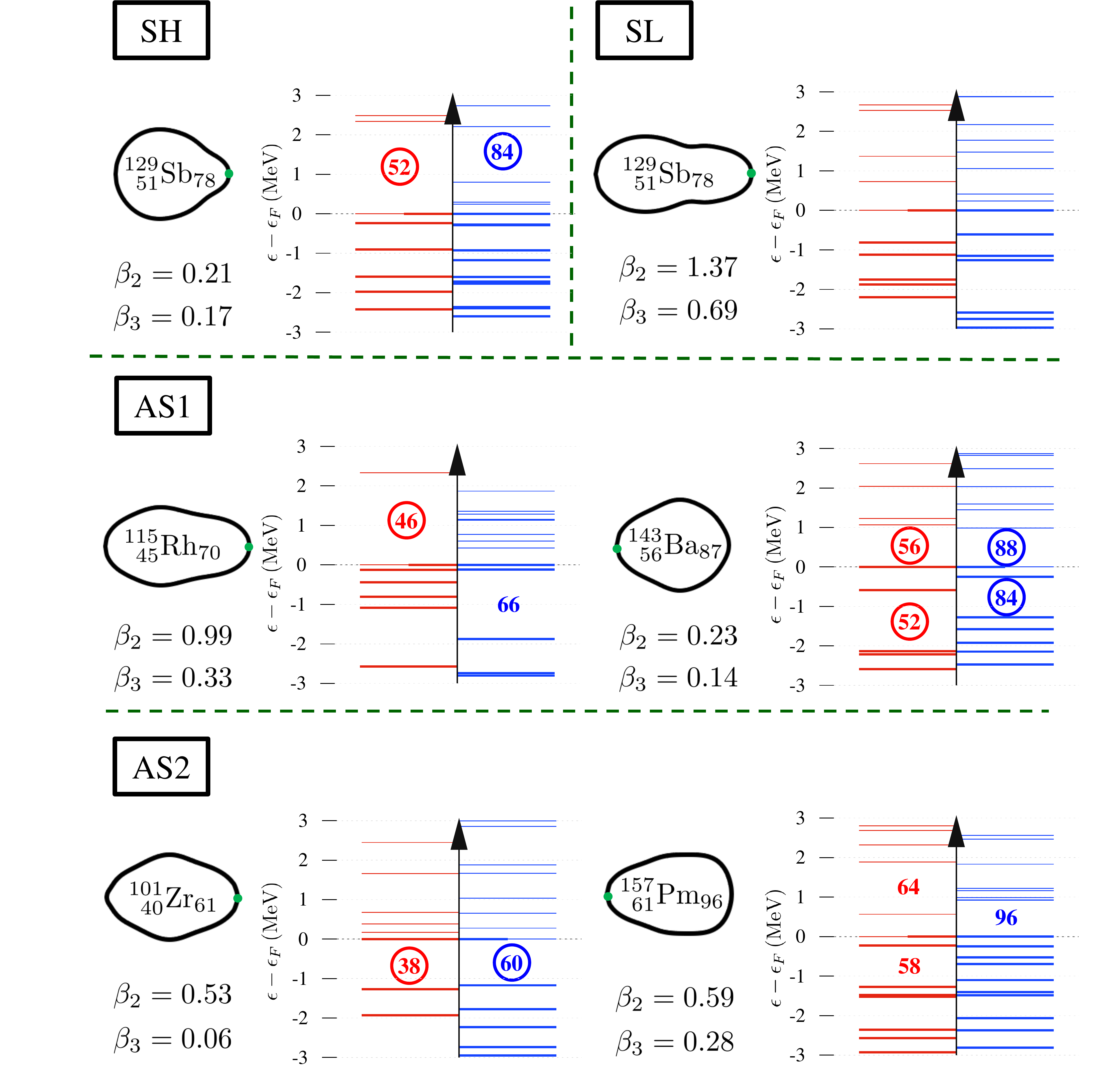}
	\caption{Representative fragment shapes with their quadrupole and octupole deformations, $\beta_2$ and $\beta_3$, and the corresponding single-particle levels for the SH, AS1, AS2, and SL modes.
		Proton (red) and neutron (blue) single-particle levels are shown to the left and right of the vertical arrows, respectively.
		Occupied and unoccupied levels are shown by thick and thin lines, respectively.
		The single-particle energies $\epsilon$ are measured relative to the Fermi energy $\epsilon_F$.
		Circles mark the particle numbers corresponding to selected level gaps discussed in the text.
		The green dots indicate the neck points of the fragment shapes.}
	\label{fig:frag_shell}
\end{figure}

To relate the fission modes to fragment shell structure, representative scission configurations at $E^*=6$~MeV are analyzed in terms of fragment shapes and single-particle levels.
Figure~\ref{fig:frag_shell} shows the calculated single-particle levels and representative fragment shapes for each mode.
The assigned proton and neutron numbers shown inside the shapes are used only to define representative fragments for the single-particle calculation, whereas the shell structure is examined in terms of level gaps near the Fermi surface of each isolated fragment.

For SH, the representative $^{129}$Sb fragment (complementary to $^{129}$Sn) has a compact shape with $\beta_2=0.21$ and $\beta_3=0.17$.
The single-particle levels show a prominent proton gap at $Z=52$ rather than at the spherical shell closure at $Z=50$, and a larger neutron gap at $N=84$ than at $N=82$.
The SH fragment is therefore characterized by deformed shell gaps similar to those associated with quadrupole and octupole deformations in the $^{144}$Ba region~\cite{scamps2018}.
This compact but deformed configuration is distinct from the SS mode, which is associated with compact fragments near the spherical $^{132}$Sn shell closure and with substantially higher TKE~\cite{hulet1986,brosa1990}.

AS1 has an asymmetric scission configuration with representative fragments $^{115}$Rh and $^{143}$Ba.
The heavy fragment is compact but deformed, with $\beta_2=0.23$ and $\beta_3=0.14$, and its single-particle levels show proton gaps at $Z=52$ and 56 and neutron gaps at $N=84$ and 88.
Thus, both SH and AS1 are characterized by a compact fragment with similar deformed shell gaps.
The distinction between the two modes is given by the complementary fragment, compact in SH or strongly elongated in AS1.

The AS1 light fragment has $\beta_2=0.99$ and $\beta_3=0.33$, and its proton and neutron single-particle levels show gaps at $Z=46$ and $N=66$, respectively.
This $Z=46$ proton shell gap is close in proton number to the $Z=44$ shell gap associated with the elongated light prefragment in the HFB study of asymmetric fission in $^{256}$Fm~\cite{bernard2023}.
A proton gap at $Z=46$ also appears at large deformation near $\beta_2\simeq1.0$ in calculations of sub-lead fission, as shown in Fig.~3 of the Supplemental Material for Ref.~\cite{scamps2019_sublead}.
In the six-dimensional Langevin study of $^{236}$U fission~\cite{Okada236U_companion}, a similar elongated shell pattern with gaps at $Z=46$ and $N=66$ is instead found in the symmetric SL configuration.
These comparisons suggest that proton shell effects at $Z=46$ play a role in the elongated configuration of the AS1 light fragment in $^{258}$Md$^*$ fission.

The representative fragments for AS2 are $^{101}$Zr and $^{157}$Pm.
AS2 has a more mass-asymmetric scission configuration than AS1, with both fragments appreciably deformed: $(\beta_2,\beta_3)=(0.53,0.06)$ for the light fragment and $(0.59,0.28)$ for the heavy fragment.
For the heavy fragment in AS2, the single-particle levels show proton gaps at $Z=58$ and 64 and a neutron gap at $N=96$.

The single-particle levels of the AS2 light fragment show gaps at $Z=38$ and $N=60$.
A proton gap at $Z=38$ is also found for the light fragment in a microscopic study of asymmetric $^{236}$U fission~\cite{bernard2023} and for the deformed light fragment of the dominant asymmetric mode in the six-dimensional Langevin study of $^{236}$U fission~\cite{Okada236U_companion}.
In $^{236}$U, the combination of the heavy fragment shell structure at $Z=56$ and $N=88$ with the light fragment shell structure at $Z=38$ characterizes the dominant asymmetric mode.
One possible interpretation is that the heavy- and light-fragment shell structures that occur together in the dominant asymmetric mode of $^{236}$U appear separately in the AS1 and AS2 modes of $^{258}$Md$^*$, respectively.

For the neighboring nucleus $^{258}$Fm, fission with strong mass asymmetry is suggested by the following predictions from other models.
The preformed cluster model, with static quadrupole deformations for both fragments, predicts the strongly mass-asymmetric split ${}^{102}_{40}\mathrm{Zr}_{62}+{}^{156}_{60}\mathrm{Nd}_{96}$~\cite{kaur2021}.
Three-dimensional Langevin calculations based on the two-center shell model and assuming the same deformation for both fragments also give an asymmetric component at $A_L/A_H\simeq103/155$~\cite{miyamoto2019,takagi2025}.
A similar asymmetric fission mode is also found in the present six-dimensional calculation as AS2, but its yield is smaller than that of AS1 with $A_L/A_H=115/143$.

SL is a symmetric elongated configuration whose representative $^{129}$Sb fragment (complementary to $^{129}$Sn) is strongly deformed, with $\beta_2=1.37$ and $\beta_3=0.69$.
Its single-particle levels show no clear shell gap comparable to those found for the other modes.
This indicates a more liquid-drop-like character at the scission point.